\documentstyle[preprint,prd,aps,floats,eqsecnum,epsf]{revtex}
\newcommand{\edc}{\end{document}}
\newcommand{\bb} {\begin{thebibliography}}
\newcommand{\eb}{\end{thebibliography}}
\newcommand{\bi}[1]{\bibitem{#1}}
\newcommand{\bc}{\begin{center}}
\newcommand{\ec}{\end{center}}

\newcommand{\be}{\begin{equation}\small}
\newcommand{\ee}{\end{equation}\normalsize}
\newcommand{\bea}{\begin{eqnarray}}
\newcommand{\eea}{\end{eqnarray}}
\newcommand{\ba}{\begin{array}{l}   }
\newcommand{\lab}[1]{\label{#1}}
\newcommand{\ea}{\end{array}}

\newcommand{\dsfrac}{\displaystyle\frac}
\newcommand{\ds} {\displaystyle}

\newcommand{\re}[1]{(\ref{#1})}

\newcommand{\ci}{\cite}

\newcommand{\dsint}{\ds\int}

\newcommand{\nwl}{\\[1mm]}

\newcommand{\hash}{\hbar }
\newcommand{\hc}{\hash\,c }

\newcommand{\ejdj}    {\ds{\exp\{  jDj/2 \} } }
\newcommand{\ejgj} {\ds{\exp{\{j_AGj_A/2 \}}} }
\newcommand{\ejdjnol} {\ds{\exp\{  jD_0j/2 \} } }
\newcommand{\ejgjnol} {\ds{\exp\{j_AG_0j_A/2 \}} }

\newcommand{\intdf} {\ds{\int  \cal{D} \phi } }

\newcommand{\veff}{V_{\mbox{eff}}}
\newcommand{\hint}{H_{\mbox{int}}}

\newcommand{\cl}{  \ell     }
\newcommand{\dd}{\partial}

\newcommand{\half}{\frac{1}{2}}
\newcommand{\mnol}{m_{0}^{2}}

\newcommand{\omnol}{\Omega_{0}^{2}}
\newcommand{\delnol}{\Delta_{0}^{2}}
\newcommand{\om}{\Omega^2}
\newcommand{\del}{\Delta^2}
\newcommand{\omb}{{\bar{\Omega}}\,}
\newcommand{\delb}{{\bar{\Delta}}\,}

\newcommand{\inolm}{I_0(\Omega)}
\newcommand{\inold}{I_0(\Delta)}

\newcommand{\hata}[1]{\hat{A}^{(#1)} }
\newcommand{\hatb}[1]{\hat{B}^{(#1)} }

\newcommand{\phiz}{\phi_0}

\newcommand{\tochka}{\, .}
\newcommand{{\vergul}}{\, ,}
\newcommand{\vecA}{\vec{A}}
\newcommand{\phizb}{\bar{\phi}_0}
\newcommand{\mc}{m_{c}^{2} }
\newcommand{\veps}{\varepsilon }

\begin{document}
\draft
\title{\Large\bf {  Post  Gaussian effective potential in the Ginzburg Landau
theory of superconductivity}\\
}
\author{
    Chul  Koo Kim\thanks{E-mail: ckkim@phya.yonsei.ac.kr}, A. Rakhimov \thanks
{E-mail: rakhimov@rakhimov.ccc.uz. Permanant address: Institute of Nuclear Physics, Tashkent,
    Uzbekistan (CIS)}  and
 Jae Hyung Yee \thanks{E-mail: jhyee@phya.yonsei.ac.kr}
 }
\address{
Institute of Physics and Applied Physics, Younsei University, Seoul, 120-749, Korea\\}
\maketitle
\begin{abstract}
\medskip
\medskip
\medskip

The post  Gaussian effective potential in $D=3$ dimensions and
the Gaussian effective potential  in $D=2+\veps$ are evaluated
 for the Ginzburg-Landau
 theory  of superconductivity.
 It is shown that,
the next order correction  to the Gaussian approximation of the Ginzburg-Landau
parameter $\kappa$  is significant,
whereas contribution from the two dimensionality is rather small.
This strongly indicates that strong correlation plays a more  dominant role
than the two dimensionality does in  high $T_c$ superconductivity.

\end{abstract}
\medskip
\medskip

\newpage

\section{Introduction}

 The Ginzburg-Landau (GL) theory of  superconductivity \ci{kl1} was proposed long
 before the famous BCS microscopic theory of superconductivity was discovered.
A few years after the appearance of the BCS theory,
Gorkov derived \ci{Gorkov} the GL theory from the  BCS
theory. Amazingly , the GL theory has played a significant role
in understanding superconductivity up to now. 
It is highly relevant for the description of high- $T_c$ superconductors,
even though the original BCS theory is inadequate to treat these materials.
The success  of the GL theory in the study of modern problems of superconductivity
lies on its universal effective  character  in which 
  the details of the microscopic model are  
unimportant. 

Even in the level of  meanfield approximation (MFA), the GL theory gives
significant information  
such as penetration depth ($l$)  and coherence length ($\xi$) of the
superconducting  samples.
Many unconventional properties of superconductivity connected with
the break down of the simple MFA has been studied both analytically 
 \ci{cam1}
and numerically using the GL theory\ci{cam3}.
 Particularly, the fluctuations of the gauge field
 were studied recently by Camarda et. al. \ci{camarda}
and Abreu et. al. \ci{abreu} in the Gaussian approximation of
the  field theory. The effective mass parameters of the 
Gaussian effective potential (GEP),
$\Omega$ and $\Delta$ , were interpreted as inverses of the coherence length 
$\xi=1/\Omega$ and of the  penetration depth $\cl=1/\Delta$, respectively.

In this note, we take one step further estimating corrections
to the Gaussian effective potential for the $U(1)$ scalar electrodynamics
where it represents the standard static GL effective model of superconductivity.
Although it was found that, in the  covariant pure $\lambda\phi^4$ theory in $3+1$
dimensions,   corrections  to the GEP are not large  \ci{stev}, we do not expect
them to be negligible in three dimensions for high
 $T_c$ superconductivity, where the system is strongly
correlated.

Apart from the strong correlation , another important factor,
which one should consider for high $T_c$ superconductivity, 
is the dimensionality of the system. It is well known that, most of the
high $T_c$ superconducting materials have layered structures,
which strongly suggests two-dimensional nature of high $T_c$
superconductivity. In order to test relative importance 
of the dimensionality contribution compared
to the  post Gaussian corrections, we shall also study the case of fractal dimension, 
 $D=2+\varepsilon$.

The paper is organized as follows: in Section II the GL  action is
introduced and the post Gaussian approximation  is applied; in Section 
III, the theoretical results for $D=3 $
and $D=2+\veps$ will be compared 
to existing high $T_c$ experimental data.
 The results are summarised in Section IV.

\section{Post Gaussian effective potential in D=3 dimension}

We start with the Hamiltonian of the GL model in Euclidean D-dimensional space
given by \ci{kleinertbook}
\bea
\nonumber
{\cal{H}'}&=&\dsfrac{1}{T_c}\dsint d^Dx\{\dsfrac{1}{4}F_{ij}^{2}+
\half\vert (\dd_i-i e \mu^{(3-D)/2}A_i)\psi\vert^2
\nwl
&  +&\half m^2\psi^2+\lambda\mu^{(3-D)}\vert\psi\vert^4\}\vergul
\lab{hprim}
\eea
where $\psi$ and $\vecA$ are the complex scalar and the static electromagnetic fields, 
respectively; $m$, $\lambda$ and $e$ are the input parameters of the model.
\footnote{$\mu$  is introduced to make $\lambda$ and $e$ dimensionless.} 
We introduce natural units  employing $\xi_0$ (coherence length at zero temperature)
and $T_c$ as length and energy scale, respectively, through the transformations : 
\be
\ba
m\rightarrow{m}\xi_{0}^{-1},\quad
 \mu\rightarrow{\mu}\xi_{0}^{-1},\quad  x\rightarrow{x}\eta_0, 
\nwl
e^2\rightarrow{e^2}\xi_{0}^{-1}T_{c}^{-1}\, ,\quad
 \lambda\rightarrow{\lambda}\xi_{0}^{-1}T_{c}^{-1}\vergul
\nwl
\psi\rightarrow{\psi}\xi_{0}^{(1-D/2)}T_{c}^{1/2}\vergul \quad
A_i\rightarrow{A_i}\xi_{0}^{(1-D/2)}T_{c}^{1/2} \tochka
\lab{scale}
\ea
\ee
Eq.  \re{hprim} is now  rewritten as,
\bea
\nonumber
{\cal{H}'}&=&\dsint d^Dx\{\half\vert \vec{\nabla}\times\vecA\vert^2+
\half\vert (\dd_i-i e \mu^{(3-D)/2}A_i)\psi\vert^2
\nwl
&+&\half m^2\psi^2+\lambda\mu^{(3-D)}\vert\psi\vert^4\}\tochka
\lab{hp}
\eea
In accordance with refs.  \ci{abreu,camarda},  we apply tranverse unitary gauge and express
 the partition function as
\be
Z= \intdf{\cal{D}}A_T\exp\{-\ds\int d^Dx H+\ds\int d^Dx j\phi
+(\vec{j}_A \vecA)\}
\lab{Z}
\ee
where the Hamiltonian density is  \footnote{From now on,
 we denote
$\lambda\mu^{(3-D)}$ and  $e^2\mu^{(3-D)}$ as $\lambda$ and $e^2$, 
respectively,  for simplicity.}
\bea
\nonumber
H&=&\half( \vec{\nabla}\times\vecA)^2+\half( \vec{\nabla}\phi)^2
+\half m^2\phi^2+\lambda\phi^4   +\half e^2\phi^2 A^2
\nwl
&+&\dsfrac{1}{2\varepsilon} ( \vec{\nabla}\vecA)^2\tochka
\lab{h}
\eea
We have introduced  a gauge fixing term, with the limit
$\varepsilon\rightarrow 0$ being taken 
after the calculations are carried out.
In Eq. \re{h} $\vecA $ stands for the transverse gauge field 
and $\psi=\phi\exp(i\gamma) \tochka$ 

To obtain the free energy density, $\veff={\cal{F}}/{\cal{V}}$
(effective potential), we introduce a shifted field 
$\phi\rightarrow\phi+\phi_0$ and split the Hamiltonian into two parts: 
\be
H=H_0+\hint \vergul
\ee
where $H_0$ is the sum of two free  field terms describing a vector
field $\vecA$ with mass $\Delta_0$ and a real scalar field $\phi$
with mass $\Omega_0$:
\bea
\nonumber
 H_0&=&\half( \vec{\nabla}\times\vecA)^2+\half \bigtriangleup^2\vecA^2+ 
\dsfrac{1}{2\varepsilon} ( \vec{\nabla}\vecA)^2
\nwl
&+&\half( \vec{\nabla}\phi)^2+\half \Omega_0^2\phi^2\tochka
\lab{h0}
\eea
The interaction term then reads
\be
\ba
\hint(\phi,A)=\ds\sum_{n=0}^{4}v_n\phi^n-\half\delnol
\vecA^2+\half e^2\vecA^2(\phi+\phi_0)^2\vergul
\lab{hint}
\ea
\ee
 where
\be
\ba
v_0=\half m^2\phiz^2+\lambda\mu^{(3-D)}\phiz^4\vergul
 \quad 
v_1=m^2\phiz+4\lambda\mu^{(3-D)}\phiz^3 \vergul
\nwl
v_2=\half (m^2-\omnol)+6\lambda\mu^{(3-D)}\phiz^2 \vergul\quad 
v_3= 4\lambda\mu^{(3-D)}\phiz,
\nwl 
v_4=\lambda\mu^{(3-D)}\tochka
\lab{v04}
\ea
\ee
Now performing explicit Gaussian integration in Eq. \re{Z},  one obtains
\small
\bea
\nonumber
Z&=&\ds \exp\{-\int d^Dx \hint (\delta/\delta j, \delta/\delta j_A  )\}
\intdf  {\cal{D}}A     \ds  \exp\{
\nwl
\nonumber
&-&\int d^Dx H_0+j\phi+\vec{j}_A\vecA ) \}
\nwl
\nonumber
&=&[\det D_{0}^{-1}]^{-\half}[\det G_{0}^{-1}]^{-\half}
\ds \exp\{-\int d^Dx \hint (\delta/\delta j, \delta/\delta j_A  )\}
\nwl
&\times&
\ejdjnol \ejgjnol {\vergul}
\lab{z2}
\eea
\normalsize
where in momentum space
\be
D_0(p)=1/(p^2+\omnol),  \quad G_0(p)=2/(p^2+\delnol)    \tochka
\lab{d0g0}
\ee
To calculate partition function  in post Gaussian approximation,
 we use the method
introduced in refs. \ci{oursingap,yee}      
    and introduce the so called primed derivatives:
\small
\be
\ba
(\dsfrac{\delta}{\delta j(x)})'\equiv\hata{1}_x
=\dsfrac{\delta}{\delta j(x)}{\vergul}\quad
(\dsfrac{\delta}{\delta j_{A}(x)})'\equiv\hatb{1}_x
=\dsfrac{\delta}{\delta j_A(x)}{\vergul}
\nwl
\\
(\dsfrac{\delta^2}{\delta j^{2}(x)})'\equiv\hata{2}_x
=\dsfrac{\delta^2}{\delta j^{2}(x)}-D_0(x{\vergul}x){\vergul}
\\
\nwl
(\dsfrac{\delta^2}{\delta j_{A}^{2}(x)})'\equiv\hatb{2}_x
=\dsfrac{\delta^2}{\delta j^{2}(x)}-G_0(x{\vergul}x){\vergul}
\\
\nwl
(\dsfrac{\delta^3}{\delta j^{3}{(x)}})'\equiv\hata{3}_x
=\dsfrac{\delta^3}{\delta j^{3}{(x)}}-3 D_0(x{\vergul}x)    R(x){\vergul}
\nwl
\\
(\dsfrac{\delta^3}{\delta j_{A}^{3}{(x)}})'\equiv\hatb{3}_x
=\dsfrac{\delta^3}{\delta j_{A}^{3}{(x)}}-3 G_0(x{\vergul}x)    R_A(x){\vergul}
\nwl
\\
(\dsfrac{\delta^4}{\delta j^{4}{(x)}})'\equiv \hata{4}_x
=\dsfrac{\delta^4}{\delta j^{4}{(x)}}-
6D_0(x{\vergul}x)\dsfrac{\delta^2}{\delta j^{2}{(x)}} +     3D_{0}^2(x{\vergul}x) {\vergul}
\nwl
\\
(\dsfrac{\delta^4}{\delta j_{A}^{4}{(x)}})'\equiv \hatb{4}_x
=\dsfrac{\delta^4}{\delta j_{A}^{4}{(x)}}-
6G_0(x{\vergul}x)
\dsfrac{\delta^2}{\delta j_{A}^{2}{(x)}} +   3G_{0}^2(x{\vergul}x)\vergul
\lab{primder}
\ea
\ee
\normalsize
where $R(x)=\ds\int d^Dy D_0(x{\vergul}y)j(y)$
 and $R_A(x)=\ds\int d^Dy G_0(x{\vergul}y)j_A(y)$, so that
\be
\ba
\hata{n}_{x}\ejdjnol= R^{n}(x)\ejdjnol{\vergul}  
\nwl
\hatb{n}_{x}\ejgjnol= R_{A}^{n}(x)\ejgjnol\tochka
\lab{bfm3.9}
\ea
\ee
Now it can be shown that \ci{oursingap,yee},  the Gaussian part of 
$Z$ can  easily be isolated  as follows:
\bea
\nonumber
Z&=&Z_{G}\;\Delta Z\vergul
\nwl
\nonumber
Z_{G}&=&\exp\{-I_1(\Omega)-\half  I_1(\Delta)-v_0-v_2\inolm+3v_4I_{0}^{2}
(\Omega) 
\nwl
\nonumber
&+&(\delnol+e^2\inolm-e^2\phiz^2)\inold    \}\vergul
\nwl
\nonumber
\Delta Z&=&\exp\{-v_2\hata{2}-v_3\hata{3}-v_4\hata{4}
\nwl
&-&\half (e^2\phiz^2-\delnol)\hatb{2}-e^2\phiz\hatb{2}\hata{1}
\nwl
\nonumber
&-&\half e^2\hatb{2}\hata{2}  \} \ejdj\ejgj\vergul
\lab{zz}
\eea        
where
\be
\ba
D(p)=1/(p^2+\om),  \quad G(p)=2/(p^2+\del),
\nwl
\om=\omnol+12v_4\inolm+2e^2\inold,\quad \del=\delnol+e^2\inolm\tochka
\lab{eqom}
\ea
\ee
In the above, following integrals are introduced
\be
\ba
I_0(M)=\dsint\dsfrac{d^Dp}{(2\pi)^D}\dsfrac{1}{(M^2+p^2)},
 \nwl
 I_1(M)=\dsfrac{1}{2}\dsint\dsfrac{d^Dp}{(2\pi)^D}\ln(M^2+p^2)\tochka
\lab{i0i1}
\ea
\ee
From Eqs. (2.14),
\re{eqom} and \re{v04}, one gets the following  Gaussian effective potential:
\bea
\nonumber
V_G&=&-\ln Z_G=I_1(\Omega)+\half  I_1(\Delta)+
v_0+v_2\inolm
\nwl
\nonumber
&-&3v_4I_{0}^{2}(\Omega) -(\delnol+e^2\inolm-e^2\phiz^2)\inold
\nwl
&=&I_1(\Omega)+\half  I_1(\Delta)+\half m^2 \phiz^2+\lambda\phiz^4
\nwl
\nonumber
&+&\half \inolm[m^2-\om+6\lambda\inolm+12\lambda\phiz^2]
\nwl
\nonumber
&+&\inold[-\delnol+e^2\inolm+e^2\phiz^2]\tochka
\lab{vg}
\eea
Note that, the last equation is exactly the same as it is in refs. \ci{camarda,abreu}.
The post Gaussian effective potential 
\be
\veff=V_G+\Delta V_G
\ee
includes a correction part $\Delta V_G$ :
\bea
\nonumber
\Delta V_G&=&-\ds\ln \Delta Z =-\ds\ln\{  \exp[-\delta\hat{W}]
\nwl
\nonumber
&\times& \ejdj\ejgj \vert_{j=0,j_A=0}  \}
\nwl
\nonumber
&=&-\ln\{1-\delta\hat{W}  \ejdj\ejgj \vert_{j=0,j_A=0}
\nwl
\nonumber
&+&\dsfrac{\delta^2{\hat{W}}^2}{2!}
 \ejdj\ejgj \vert_{j=0,j_A=0}+\dots)
    \}
\nwl
&\equiv &\delta\Delta V_{G}^{(1)} (B)+
\delta^2\Delta V_{G}^{(2)} (B)+\dots  {\vergul}
\nwl
\nonumber
\hat{W}&=&v_2\hata{2}+v_3\hata{3}
+v_4\hata{4}
+\half (e^2\phiz^2-\delnol)\hatb{2}
\nwl
\nonumber
&+&e^2\phiz\hatb{2}\dsfrac{\delta}{\delta j}
+\half e^2\hatb{2}\hata{2}\tochka
\lab{dv}
\eea
Here we have introduced an auxiliary expansion parameter $\delta$ to be
set equal to unity after calculations.

The first order term     $ \Delta V_{G}^{(1)} (B)$
in  this equation will not contribute to the effective potential,  i.e.,
 $ \Delta V_{G}^{(1)} (B)=0$,   due to the relations    \re{bfm3.9}.
 The next term  of order $\delta^2$ gives the first nontrivial contribution to the 
post Gaussian effective potential. 
The explicit calculations give
\bea
\nonumber
\Delta V_G&=& [ - {\displaystyle \frac {1}{2}} \,e^{4}\,
\mathrm{I_2}(\Delta ) - 18\,\mathrm{I_2}(\Omega )\,\lambda ^{2}
]\,{\phi _{0}}^{4}
 + \{ 
 - 3\,\lambda \,\mathrm{I_2}(\Omega )\,
\nwl
\nonumber
&\times&
[ - \Omega ^{2} + m^{2}
 + 2\,\mathrm{I_0}(\Delta )\,e^{2} + 12\,\lambda \,\mathrm{I_0}
(\Omega )]
\nwl
 &-& e^{2}\,\mathrm{I_2}(\Delta )\,[ - \Delta ^{2} + e^{
2}\,\mathrm{I_0}(\Omega )] 
\mbox{} - 8\,\lambda ^{2}\,\mathrm{I_3}(\Omega,\Omega ) 
\nwl
\nonumber
&-& {\displaystyle \frac {2}{3}} \,e^{4}\,\mathrm{I_3}(\Delta , \,
\Omega )\}{\phi _{0}}^{2}
- {\displaystyle \frac {1}{8}} \,\mathrm{I_2}(\Omega )\,
[ - \Omega ^{2} + m^{2} + 2\,\mathrm{I_0}(\Delta )\,e^{2} 
\nwl
\nonumber
&+& 12\,
\lambda \,\mathrm{I_0}(\Omega )]^{2}
 - {\displaystyle \frac {1}{
2}} \,\mathrm{I_2}(\Delta )\,[- \Delta ^{2} + e^{2}\,\mathrm{
I_0}(\Omega )]^{2}
\nwl
\nonumber
 &-& {\displaystyle \frac {1}{12}} 
\,e^{4}\,\mathrm{I_4}(\Delta , \,\Omega ) - {\displaystyle 
\frac {1}{2}} \,\lambda ^{2}\,\mathrm{I_4}(\Omega,\Omega )\vergul
\lab{vcor}
\eea
where following loop integrals were introduced,
\be
\ba
I_2(M)=\dsfrac{2}{(2\pi)^{D}}\dsint \dsfrac{d^D k}{(k^2+M^2)^2}
\nwl
\\
I_3(M_1,M_2)=\dsfrac{1}{(2\pi)^{2D}}\dsint \dsfrac{d^D k\, d^D p}
{
(k^2+M_{1}^{2})
   (p^2+M_{1}^{2}) ((k+p)^2+M_{2}^{2})         }\vergul
\\
\nwl
I_4(M_1,M_2)=\dsfrac{1}{(2\pi)^{3D}}
\dsint \dsfrac{d^D k\, d^D p\, d^D q}
{
(k^2+M_{1}^{2})
   (p^2+M_{1}^{2})(q^2+M_{2}^{2})
 }
\nwl
\\
\times\dsfrac{1}{
((k+p+q)^2+M_{2}^{2})
}         
\tochka
\lab{integs}
\ea
\ee
For $D=3-2\varepsilon$,  these integrals were calculated in
 dimensional regulariztion in ref.
\ci{integs}. For completeness 
explicit expressions are given in the Appendix. The appropriate counter terms 
to cancel the divergences coming from the integerals are also presented in the Appendix.

The parameters $\Omega$ and $\Delta $ are determined by
the  principle of minimal sensitivity
(PMS) :
\small
\bea
\nonumber
\dsfrac{\dd \veff}{\dd\omb}& =& 576\,\delb \,\pi ^{2}\,\lambda ^{2}\,\phizb ^{4} (2\,
\delb  + \omb )\,+ 8\pi \phizb^{2} \mbox{}\{ - 12\,\lambda \,e^{2}\,\delb 
^{3} 
\nwl
\nonumber
&+& [ - 72\,\pi \,\lambda \,\omb ^{2} + ( - 6\,\lambda \,
e^{2} + 192\,\lambda ^{2})\,\omb  + 24\,\pi \, {m^2}\,
\lambda ]\,\delb ^{2}
\nwl
\nonumber
& +& \delb\,[- 36\,\pi \,\lambda \,\omb ^{3} + (10\,e^{4} + 96\,
\lambda ^{2})\,\omb ^{2} 
\nwl
\nonumber
&+& 12\,\pi \,\omb \,\lambda \,
 {m^2}]\,  + \omb ^{3}\,e^{4}\}\, + 
20\,\delb \,\omb ^{5}\,\pi ^{2} 
\nwl
\nonumber
&+& (48\,\pi \,\lambda \,\delb  + 40\,\delb ^{2}\,\pi ^{2
} - 2\,e^{4})\,\omb ^{4} 
\nwl
\nonumber
&-& 4\,\delb\omb ^{3} [ - 24\,\pi \,\lambda \,
\delb  - \delb \,\pi \,e^{2} + 2\,e^{4}\,  {\ln}2 + e^{4}
\nwl
\nonumber
 &- &15\,\lambda ^{2} - 2\,e^{4}\,  {\ln}
{\displaystyle \frac {\mu ^{2}}{(\delb  + \omb )^{2}}}  + 6\,
\pi ^{2}\, {m^2} - 24\,\lambda ^{2}\,  {\ln}
{\displaystyle \frac {\mu ^{2}}{\omb ^{2}}} 
\nwl
&+& 96\,\lambda ^{2
}\,  {\ln}2 + e^{4}\,  {\ln}{\displaystyle \frac {\mu 
^{2}}{\delb ^{2}}} ]
\nwl
\nonumber
& -& 8\,\delb ^{2}\omb ^{2}\mbox{}
[ - \delb \,\pi \,e^{2} + e^{
4}\,  {\ln}{\displaystyle \frac {\mu ^{2}}{\delb ^{2}}}
 + 6\,\pi ^{2}\, {m^2}
\nwl
\nonumber
 &-& 2\,e^{4}\,  {\ln}
{\displaystyle \frac {\mu ^{2}}{(\delb  + \omb )^{2}}}  - 15
\,\lambda ^{2} + 2\,e^{4}\,  {\ln}2 
\nwl
\nonumber
&+& 96\,\lambda ^{2}\,  {\ln}2 - 24\,\lambda ^{2}\,
  {\ln}{\displaystyle \frac {\mu ^{2}}{\omb ^{2}}}]
\nwl
\nonumber
& +&\delb(2 \delb+\omb) \,(\delb \,e^{2} - 2\,\pi \, {
m^2})^{2}\,=0 ;
\nwl
\nonumber
\nwl
\dsfrac{\dd \veff}{\dd\delb}& =& 16\,\omb \,\pi ^{2}\,e^{4}\phizb ^{4}\,(\omb  + 2\,
\delb )\,
\nwl
\nonumber
&+& 8\,e^{2} \pi\phizb^{2} \,
\{( - 8\,\omb \,\pi  + 12\,\lambda )\,\delb ^{3}
\nwl
\nonumber 
&+& 2\,\omb \,\delb ^{2}( - 2\,\omb \,\pi  + 4\,e^{2} + 3\,\lambda )\,
 - e^{2}\,\omb ^{2}\,(\omb  + 6\,\delb )\}\, \,
\nwl
\nonumber
&+& e^{4}\,\omb ^{4} + 2\,\delb \,\omb ^{3}\,e^{2}\,(2\,\pi \,
\delb  + 5\,e^{2}) 
\nwl
\nonumber
 &+& 4\,\delb ^{2}\,\omb ^{2}\,[4\,\pi ^{2}\,\delb ^{2} + 2\,\pi \,e
^{2}\,\delb  + 6\,e^{4} 
\nwl
\nonumber
&+& 2\,e^{4}\,  {\ln}{\displaystyle 
\frac {\mu ^{2}}{(\delb  + \omb )^{2}}}  - 4\,e^{4}\,  {
\ln}2 - 3\,e^{2}\,\lambda ] 
\nwl
\nonumber
&+& 2 
\delb ^{2} 
[16\,\pi ^{2}\,\delb ^{3} + 7\,e^{4}\,\delb
\nwl
\nonumber
  &+& 8\,\delb \,e^{4
}\,  {\ln}{\displaystyle \frac {\mu ^{2}}{(\delb  + \omb 
)^{2}}}  - 12\,\delb \,e^{2}\,\lambda  - 16\,\delb \,e^{4}\,
  {\ln}2  
\nwl
\nonumber
&+& 2\,e^{2}\, {m^2}\,\pi ]\,\omb 
+ 4\,\delb ^{3}\,e^{2}\,( - \delb \,e^{2} + 2\,\pi \,
 {m^2}) =0\vergul
\lab{gapeq}
\eea
\normalsize
where we denote optimal values of $\Omega $
and $\Delta$ by $\omb$ and $\delb$, respectively, and 
$\phizb$ is  a stationary point defined from the equation:
\small
\bea
\nonumber
\dsfrac{\dd \veff}{\dd\phi_0}& =& 
 \{( - 4\,\pi \,e^{4} + 64\,\pi ^{2}\,\delb \,
\lambda )\,\omb  - 144\,\pi \,\lambda ^{2}\,\delb \}\,\phizb^{2}  
\nwl
\nonumber
&+& (
e^{4} - 36\,\pi \,\lambda \,\delb )\,\omb ^{2}
+2\,\omb \,
\delb [ 
 - 24\,\lambda ^{2}\,  {\ln}{\displaystyle \frac {\mu ^{2}}{
\omb ^{2}}} 
\nwl
\nonumber
 &+& 
48\,\lambda ^{2}\,  {\ln}3 - 2\,\delb \,
\pi \,e^{2} + e^{4}\,  {\ln}{\displaystyle \frac {\mu ^{2}}{
\delb ^{2}}}  - 2\,e^{4}\,  {\ln}2 
 {m^2}
 \nwl
 &+& 8\,\pi ^{2}\,- 2\,e^{4}\,  {\ln}{\displaystyle \frac {\mu ^{2}}{(
2\,\delb  + \omb )^{2}}} - 6\,\lambda ^{2}]\mbox{} 
\nwl
\nonumber
&+&
 6\,
\lambda \,\delb \,(\delb \,e^{2} - 2\,\pi \, {m^2})=0\tochka
\eea
\normalsize
Note that, in the Gaussian approximation,
 the gap equations \re{gapeq} are reduced to  simple forms:
\bea
\nonumber
\dsfrac{\dd V_G}{\dd\omb}&=&\omb ^{2}\,\pi- 6\,\omb \,\lambda  - \delb \,e^{2} + 2
{m^2}\,\pi   =0\vergul
\nwl
\nwl
\nonumber
\dsfrac{\dd V_G}{\dd\omb}&=&8\,\delb ^{2}\,\pi \,\lambda - e^{4}\,\delb  
- 4\omb \,e^{2}\,\lambda+ 2\,e^{2}\,{m^2}\,\pi 
    =0 \tochka
\lab{gapgs}
\eea

\section{Comparision with experimental data for $D=3$
and $D=2+\veps$\tochka}

The solutions of the Eqs.  \re{gapeq} are related to the experimentally measured
GL parameter $\kappa$ as $\kappa=\cl/\xi=\omb/\delb$. 
We make an attempt to reproduce  recent experimental data
on $\kappa (T)$  \ci{exper} for high -$T_c$ cuprate superconductor
$Tl_2Ca_2Ba_2Cu_3O_{10} (T\cl-2223)$. 

For  this purpose,  we adopt usual linear  $T$ dependence of parametrization
of $m$ and $\lambda$ as:
\be
\ba
m^2=\mnol(1-\tau)+\tau \mc\vergul
\nwl
\lambda=\lambda_0(1-\tau)+\tau \lambda_c\vergul
\nwl
\tau=T/T_c\vergul
\lab{param}
\ea
\ee
and calculate $\kappa$ by solving  nonlinear equations
\re{gapeq} or \re{gapgs}. Due to the parametrization \re{param},
 the model has in general
five input parameters: $\mnol$, $\lambda_0$,  $\mc$, $\lambda_c$ and $e$. 
The last   parameter is related directly to the electron charge: 
$e^2=16\pi\alpha k_B T_c \xi_0/\hc$, where $\alpha=1/137$,  $\xi_0$
is a coherence length  at $T=0$ , and $T_c$   the critical temperature.
The experimental values for the cuprate $T\cl-2223$ are $\xi_0=1.36nm$
and $T_c=121.5K$. The parameters $\mnol$ and $\lambda_0$
are fitted to the expeimental values of $\xi$ and $\cl$ at zero
temperature: $\xi_0=1.36nm$, $\cl_0=163nm$.  In dimensionless units,
\re {scale}, we have $\omb_0=\omb(\tau=0)=1$ and 
$\delb_0=\delb(\tau=0)=\xi_0/l_0=0.0083$ which are used to calculate 
$\mnol$ and $\lambda_0$ from coupled
 equations \re{gapeq} (or \re{gapgs} in the Gaussian case).
 The parameters $\mc$
and $\lambda_c$ are fixed in the similar way. Actually the quantum fluctuations
shift $\mc$ from its zero value given by MFA. On the other hand, the exact 
experimental values of $\mc$ and $\lambda_c$ are unknown, since
the GL parameter at  $T=T_c$  is poorly determined.
For this reason, we used the experimental values of $\xi_c$ and
$\ell_c$ at very close points to the critical temperature
taking $\tau_c=0.98$ which gives $\omb_c=\omb(\tau_c)=1/\xi_c=0.128$ and
$\delb_c=\delb(\tau_c)=1/\cl_c=0.0043$.  Then solving the equations \re{gapeq} 
(or \re{gapgs} in  the Gaussian case) 
with respect to $m_c$ and $
\lambda_c$, we fix   the input  parameters. Their values  for the Gaussian and
the post Gaussian cases for D=3 are summarized in Table 1  \footnote {All parameters 
are  given in dimensionless units. See Eq. \re{scale}.   }.

\begin{table} [bp]
\caption{
 Input parameters   of the GL model.
 }
\bc
\begin{tabular}{ccccc}
 & $\mnol $ &  $\lambda_0$  &   $\mc$& $\lambda_c$\\
\hline
Gaussian& -0.456 &0.046  & 0.0013 & 0.002\\
\hline
Post. Gaussian& -0.525 &0.050 & 0.0017 & 0.008\\
\end{tabular}
\ec
\end{table}

After having fixed the input parameters, the temperature dependence
of $\omb(\tau)$, $\delb(\tau)$ as well as the GL parameter $\kappa=\omb(\tau)/\delb(\tau)$
are established by solving  the gap equations 
(2.25) and \re{gapeq} numerically
for the Gaussian and the post Gaussian approximations, respectively. The results are 
presented in Fig.1 , where solid curve corresponds to the post Gaussian 
and dotted one to the Gaussian approximation. It is seen from the figure
that corrections to the Gaussian approximation are significant, and 
in the right direction, although
the discrepancy from the experimental values  is still substantial. 

On the other hand,  a better agreement with the experiment has been obtained
even on the level of the Gaussian approximation by the authors of  ref.  \ci{camarda}.
However, they introduced a cut off parameter $\Lambda$ as a characteristic 
energy scale of the sample
to make  the divergent integrals $I_0$ and $I_1$ finite. 
We beleive that, the better agreement is a result of introducing this rather 
 arbitrary additional parameter. It should be noted that, in the present
approach, there is no such additional adjustable parameter.
Here  we used dimensional regularization in which
 we put  $\mu=\omb_0$.  It was found that the behavior of 
$\kappa(\tau)$ does not depend on $\mu$:
Another value of $\mu$, e.g. $\mu=2\omb_0$
leads to another set of input parameters $\{ \mnol, m_c, \lambda_0, \lambda_0\}$,
but to the same behavior for $\kappa(\tau)$. 

Clearly, the solutions of nonlinear gap equations are not unique. In numerical calculations 
we separated  the physical  solutions by observing the sign of $\phizb^2$ and 
that of the effective potential at the stationary point : $\veff(\phizb)$. The temperature dependence
of these two quantities are presented in Fig. 2. It is seen that   $\phizb^2$ (solid line)
is positive in the large range of $\tau$ and goes to zero when $\tau$ is close to $\tau=1$.
Similarly, the depth of the effective potential at the stationary point, 
 $\veff(\phizb)$,
becomes shallow when $\tau\rightarrow1 $ and vanishes at $T=T_c$ .

All the above numerical calculations were made in D=3 dimension. On the other hand
it is widely known that, most of  high $T_c$ cuprates have
 layered structures with 2D $CuO_2$ planes 
which play an essential role  in the high $T_c $ superconductivity.
Therefore, it is nessesary to consider the dimensional contribution in the calculation
so that relative importance between the post Gaussian corrections and the two -dimensional 
character can be assessed.
For this purpose,  we consider the case of  $D=2+2\varepsilon$
$(\varepsilon\neq 0)$ in the  Gaussian approximation.
The effective potential is given by the  Eqs. 
(2.16)
 and
(2.17)
 where the integrals  are explicitely
written as:
\bea
\nonumber
 I_0(M)&=&- {\displaystyle \frac {1}{4\pi }} \,
\{
[
\dsfrac {1}{12} \,\pi ^2
 + \dsfrac {1}{2} \,{\ln^2}\dsfrac {\mu ^2}{M ^2}      
]\,\varepsilon 
 - \ln\dsfrac {\mu ^2}{M ^2} + 
\dsfrac {1}{\varepsilon }
\}\vergul
\nwl
\\
\nonumber
I_1(M)&=&  - \dsfrac {M^2}{8\pi }
\{
[
\dsfrac {1}{12} \,\pi ^2
 + \dsfrac {1}{2} \,\ln^{2}\, \dsfrac {\mu ^2}{M ^2}
 +\,\ln
\dsfrac {\mu ^2} {M ^2}  + 1
]\,\varepsilon
\nwl
\nonumber
& -& 1 -\ln\dsfrac {\mu ^2} {M ^2}
 + \dsfrac {1}{\varepsilon }
\}
\tochka
\lab{i02dim}
\eea
Taking derivatives of $V_G$ and using 
(3.2)
 leads to the following
 gap equations:
\bea
\nonumber
\dsfrac{\dd V_G}{\dd\Omega}\vert_{(\dd V_G/\dd\phiz)=0       }&=&
 6\,\lambda  + e^{2} - \varepsilon \,
[
\pi \,
\omb ^{2} + 2\,\pi \,m^{2} 
\nwl
\nonumber
&+& \,(e^{2} + 12\,\lambda )
\ln\displaystyle \frac {\mu ^2}{\omb ^2}
 +\,e^{2}{\ln}
\displaystyle \frac {\mu ^2}{\delb ^2}
]
\nwl
&+&O(\veps^2)=0
\vergul
\nwl
\nonumber
\nwl
\nonumber
\dsfrac{\dd V_G}{\dd\Delta}\vert_{(\dd V_G/\dd\phiz)=0   }&=&
4 e^{2} \lambda  + e^{4} - 2 \varepsilon  
[
4 \delb ^{2} \pi  \lambda  + e^{2} m^{2} \pi  
\nwl
\nonumber
&+&  e^{2} (e^{2} + 2 \lambda )
{\ln}\displaystyle \frac {\mu ^{2}} {\delb ^{2}}
 \nwl
&+& 2 \lambda  e^{2} \ln\displaystyle \frac {\mu ^2}{\omb ^2} 
]+O(\veps^2)=0\tochka
\lab{gap2}
\eea

The  parameters $\mnol$,  $\lambda_0$, $m_{c}^{2}$
 and $\lambda_c$ in Eqs.  (3.1) - (3.4) 
 were  adjusted to their experimental values
  in the same way  as in the previous case. As a result
we  obtain:
\be
\ba
\mnol=  {\displaystyle \frac {  1.106- 
22.827\,\varepsilon }{1.106 + 36.443\,\varepsilon
 }} ,
\quad
\lambda_ 0={\displaystyle \frac { 0.869\,\varepsilon }{
1.106 + 36.443\,\varepsilon }} \vergul
\nwl
\nwl
m_{c}^{2}= \dsfrac{ - 511.839\,\varepsilon ^{4} + 15.368\,
\varepsilon ^{3} + 0.708\,\varepsilon ^{2} - 0.00645
\,\varepsilon  
\mbox{} + 0.000016}
{
140.569\,\varepsilon ^{2} + 2982.92\,\varepsilon ^{3} + 
1.9283\,\varepsilon  + 14677.46\,\varepsilon ^{4} }\vergul
\nwl
\nwl
\lambda_c = {\displaystyle \frac {0.202\,
\varepsilon ^{3} + 0.0042\,\varepsilon ^{2} - 
0.00005\,\varepsilon }{1.405\,\varepsilon  + 
29.829\,\varepsilon ^{2} + 0.0192 + 146.77\,
\varepsilon ^{3}}}\tochka
\lab{mas2dim}
\ea
\ee

In Fig.3 and Fig.4,  we present $m^2$ and $\lambda$ vs.  $\veps$,
respectively, given by Eqs. \re{param} and \re{mas2dim}. One notes that 
for small values of $\veps$ ($0<\veps\le 0.048$)
$m^2$ becomes positive . Bearing in mind that, in the GL model the phase transition 
occurs where $m^2$ changes sign (or more exactly 
the superconductive phase holds only
for $m^2<0$), it shows that,  in the present approximation scheme,
there is no phase transition
in $D=2+2\veps$ dimension for  very  small values of $\veps$. 
This smallness of the $\veps$ value  indicates reliability
of the present post Gaussian approximation. 
Note that , $\lambda$ remains positive 
on the whole range of $\veps$ (Fig.4).

The GL parameter $\kappa(\tau)=\omb(\tau)/\delb(\tau)$ given by Eqs. 
\re{param} -  \re{mas2dim}  in $D=2+2\veps$ case ($\veps=0.1$)
is plotted in Fig.1  (dotted line). 
We find that, surprisingly, the two
curves from the Gaussian approximation almost coincide, thus indicating
that the dimensionality contribution is not significant as far as the GL 
parameter is concerned.

%

\section{Summary}

In the present article we have carried out calculations on
the Ginzburg-Landau effective potential beyond the
Gaussian approximation.
The result is used to obtain the Ginzburg-Landau parameter,
 $\kappa$, and compared with existing high $T_c$ superconductivity data.
It was shown that the post Gaussian correction which
is believed to originate from
strong correlation is substantial. In order to estimate  the
contribution from the two dimensionality of
high  $T_c$ superconducting materials, we have carried out calculations 
for $D=2+2\veps$ in the Gaussian approximation.
The result shows that the dimensionality correction
to  the three dimensional Gaussian result is rather small,
although there remains
possibility that a post Gaussian correction at $D=2+2\veps$ is much larger
than that at $D=3$, thus making the theory closer to experiment. This remains as 
a future study.

\section*{Acknowledgments}
  A.M.R.            is
indebted to the Yonsei  University  for hospitality
 during his stay, where   the main part of
this work was  performed. This research was in part
 supported by BK21 project and in part
 by Korea Research Foundation under project
numbers KRF-2003-005-C00010 and KRF-2003-005-C00011.

\newpage
\def\theequation{A.\arabic{equation}}
\setcounter{equation}{0}
\bc
{\Large\bf Appendix }
\ec
\indent
\bc
{\bf A. Explicit expression for divergent integrals.}
\ec
Here, we bring explicit expressions for the divergent integrals defined as:
\bea
\nonumber
I_0(M)&=&  - {\displaystyle \frac {M}{4\pi}} 
[1 + \varepsilon(2 + \ln\,\dsfrac {\mu ^2} {4M^2} )\,+O(\varepsilon^2)]
\vergul
\nwl
\nonumber
\nwl
\nonumber
I_1(M)&=&  - {\dsfrac {M^{3}}{36{\pi }}} \,{
 {\,[3 + \varepsilon \,(8 + 3\, {\ln}
{} \,{\displaystyle \frac {\mu ^{2}}{{4}M
^{2}}} )+O(\varepsilon^2)]}} 
\vergul
\\
\nonumber
\nwl
\nonumber
{I_2(M)}&=& {\displaystyle \frac {1}{4{\pi \,M}}} \,{
 [{1 + \varepsilon \, {\ln}{\displaystyle 
} \,{\displaystyle \frac {\mu ^{2}}{{4}M^{2}}}+O(\varepsilon^2)]}}
\vergul
\\ 
\nwl
\nonumber
{I_3(M_1,M_2)}& =& {\displaystyle \frac {1}{64{\pi ^{2}}   }} \,{\displaystyle 
\{{2\, {\ln}{ \,
{\displaystyle \frac {\mu ^{2}}{(2\,{M_{1}} + {M_{2}})^{2}}} 
 + 2 + 
{\displaystyle \frac {1}{\varepsilon }} }} +O(\varepsilon)\} }
\vergul
\\
\nonumber
\nwl
\nonumber
{I_4(M_1,M_2)}&=& -{\displaystyle \frac {1}{128\pi ^{3}}}\{ {M_{1}}\,[2\,
 {\ln}({\displaystyle \frac {\mu ^{2}}{({M_{1}} + {M_{2}})^{
2}}} ) + 8 
\nwl
\nonumber
&+& {\ln}{\displaystyle \frac {\mu ^{2}}{{M_{1}}
^{2}}}  - 6\, {\ln}2 + {\displaystyle \frac {1}{
\varepsilon }}+O(\varepsilon) ]
 \\
\nonumber
&+&M_1 \leftrightarrow 
\, M_2\}
\tochka
\lab{integseps}
\eea

In D=3 dimension the integrals $I_3 $ and $I_4$ are divergent. Following
 counter terms were introduced 
to the Hamiltonian \ci{oursingap,chiku,banerji}:
\bea
H_{cont}&=&B\phi^2m^2/2+C\lambda \phi^4+De^2\phi^2 \vecA^2
\nwl
\nonumber
&+&E\vecA^2\Delta^2(1-\delta)/2
\vergul
\eea
where in the minimal subtraction ($\overline{MS}$)  scheme  
\be
\ba
B=B_1\delta+B_2\delta^2+\dots \vergul\quad C=C_1\delta+C_2\delta^2+\dots
\vergul
\nwl
D=D_1\delta+D_2\delta^2+\dots  \vergul\quad  E=E_1\delta+E_2\delta^2+\dots
\vergul
\nwl
B_1=0,\quad C_1=0,\quad D_1=0,\quad E_1=0\quad 
\vergul
\nwl
B_2=\dsfrac{3\lambda^2}{2\pi^2\varepsilon}\vergul\quad C_2=0,\quad
 D_2=-\dsfrac{e^2}{4\pi\varepsilon}\vergul\quad
E_2=\dsfrac{e^4}{16\pi^2\varepsilon}\quad\tochka
\ea
\ee

 \bb{99}
\bi{kl1} V. L. Ginzburg and L.D. Landau, Zh. Eksp. Teor. Fiz.
{\bf 20}, 1064, (1950) .
\bi{Gorkov}L.P. Gorkov, Sov. Phys. JETP {\bf 7}, 505 (1958);
L.P. Gorkov, Sov. Phys. JETP {\bf 9}, 1364 (1959).
\bi{cam1} Z. Tesanovich, Phys. Rev. B{\bf 59}, 6449, (1999) (and references there in).
\bi{cam3} A. K. Nguyen and A. Sudbo, Phys. Rev. B{\bf 60} 15307 (1999).
\bi{camarda}  M. Camarda, G.G.N. Angilella, R. Pucci
 and F. Siringo ,  Eur.Phys.J. {\bf B33} , 273, (2003).
\bi{abreu} 
 L.M. Abreu, A.P.C. Malbouisson and I. Roditi 
 "GAUGE FLUCTUATIONS IN SUPERCONDUCTING FILMS." 
 cond-mat/0305366.
\bi{stev} I. Stancu and P. M. Stevenson,  Phys. Rev. D{\bf 42}, 2710, (1990). 
\bi{kleinertbook}H. Kleinert, {\it Gauge Fields in Condensed Matter}, Vol. 1:
Superflow and Vortex Lines,  World Scientific , Singapore,1989.
\bibitem{oursingap} A. Rakhimov  and J.H. Yee,  Int. J. Mod. Phys. {\bf A19}, 1589 ( 2004).
\bibitem{yee} G.H. Lee and J.H. Yee,  Phys. Rev. D{\bf 56}, 6573,  (1997).
\bi{integs} E. Braaten and A. Nieto, Phys. Rev. D{\bf 51, 6990, (1995)};
 A. I. Davydychev and M. Yu. 
Kalmykov Nucl. Phys. {\bf  B605}, 266, (2001); 
A. K. Rajantie Nucl. Phys. {\bf  B480}, 729, (1996).
\bi{exper} G. Brandstatter, F.M. Sauerzopf, H. W. Weber, F. Ladenberger
and E. Schwarzmann, Physica C, {\bf 235}, 1845, (1994);
\nwl
G. Brandstatter, F.M. Sauerzopf and  H. W. Weber, Phys. Rev. B{\bf 55}, 11693, (1997).
\bibitem{chiku} S. Chiku and T. Hatsuda  Phys. Rev. D{\bf 58}, 076001, (1998).
\bibitem{banerji} N. Banerjee and S. Malik Phys. Rev. D{\bf 43},3368, (1991).
  \eb
\newpage
 \begin{figure}
 \epsfxsize=12.cm
\begin{center}
\epsffile{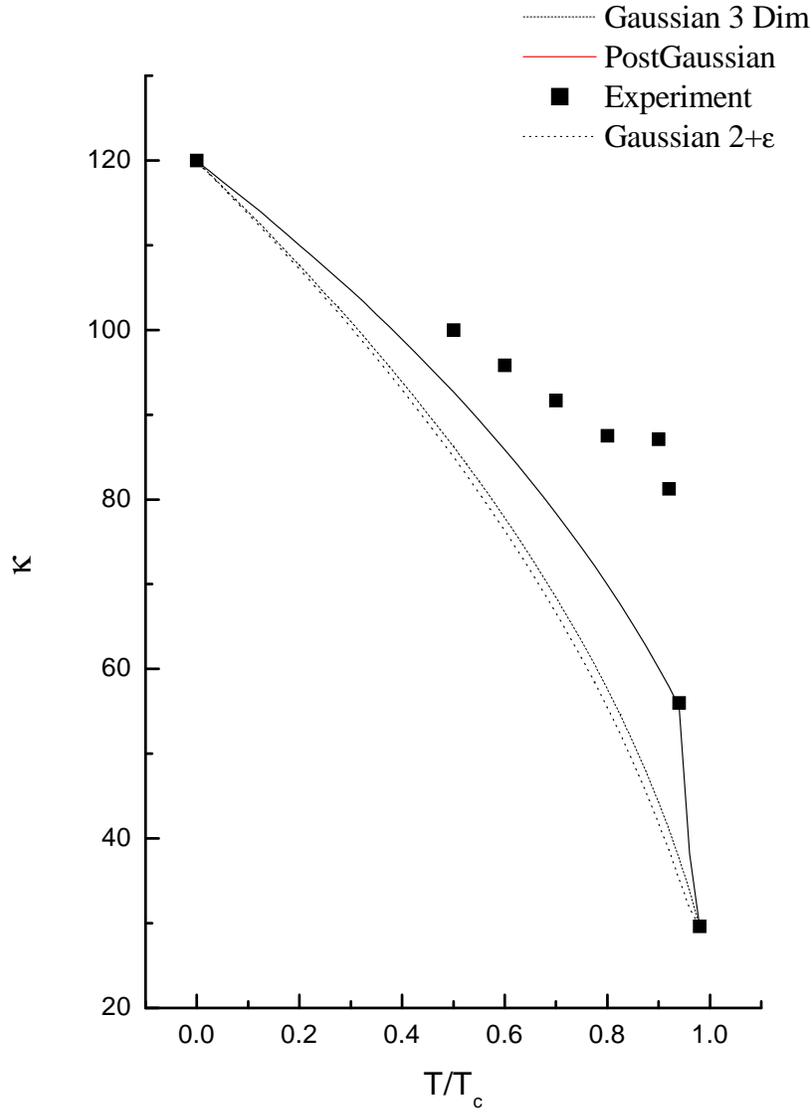}
 \end{center}
\caption{
The GL parameter, $\kappa  $, in the Gaussian approximation in 
$D=3$ (the dotted line)  and $D=2+\veps$ (the dashed line) cases. The
solid    line represents the PostGaussian approximation for  $D=3$ case.
}
\end{figure}
\newpage 
 \begin{figure}
 \epsfxsize=12.cm
\begin{center}
\epsffile{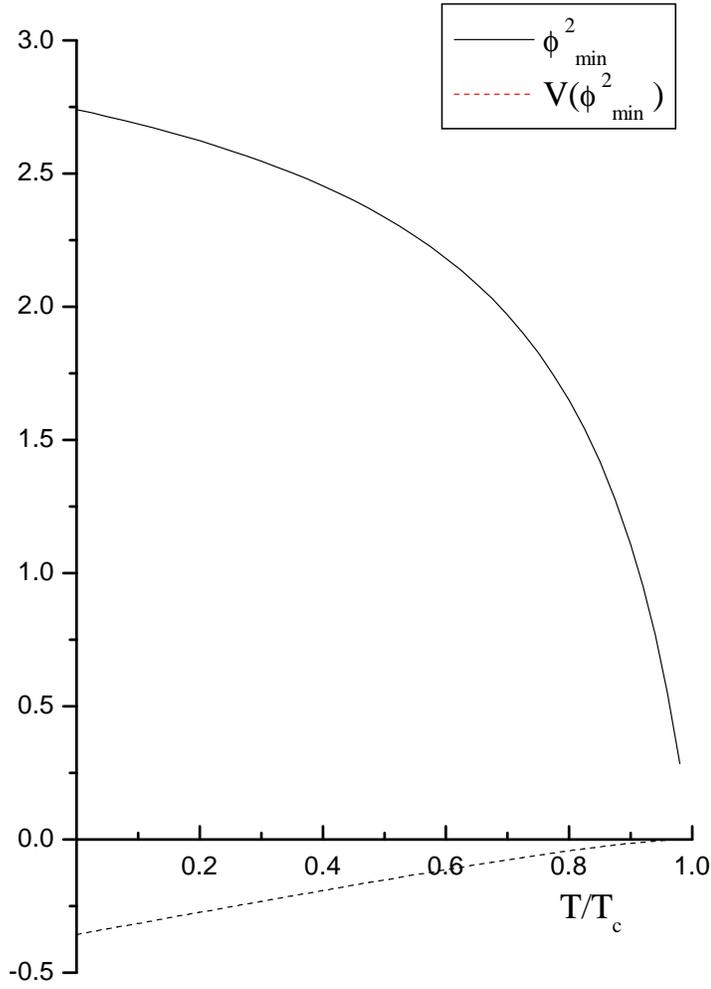}
 \end{center}
\caption{
The stationary point $\phizb^2$ and the depth of the  post Gaussian 
effective potential at the stationary  point vs. temperature.
}
\end{figure}

\newpage 
 \begin{figure}
 \epsfxsize=12.cm
\begin{center}
\epsffile{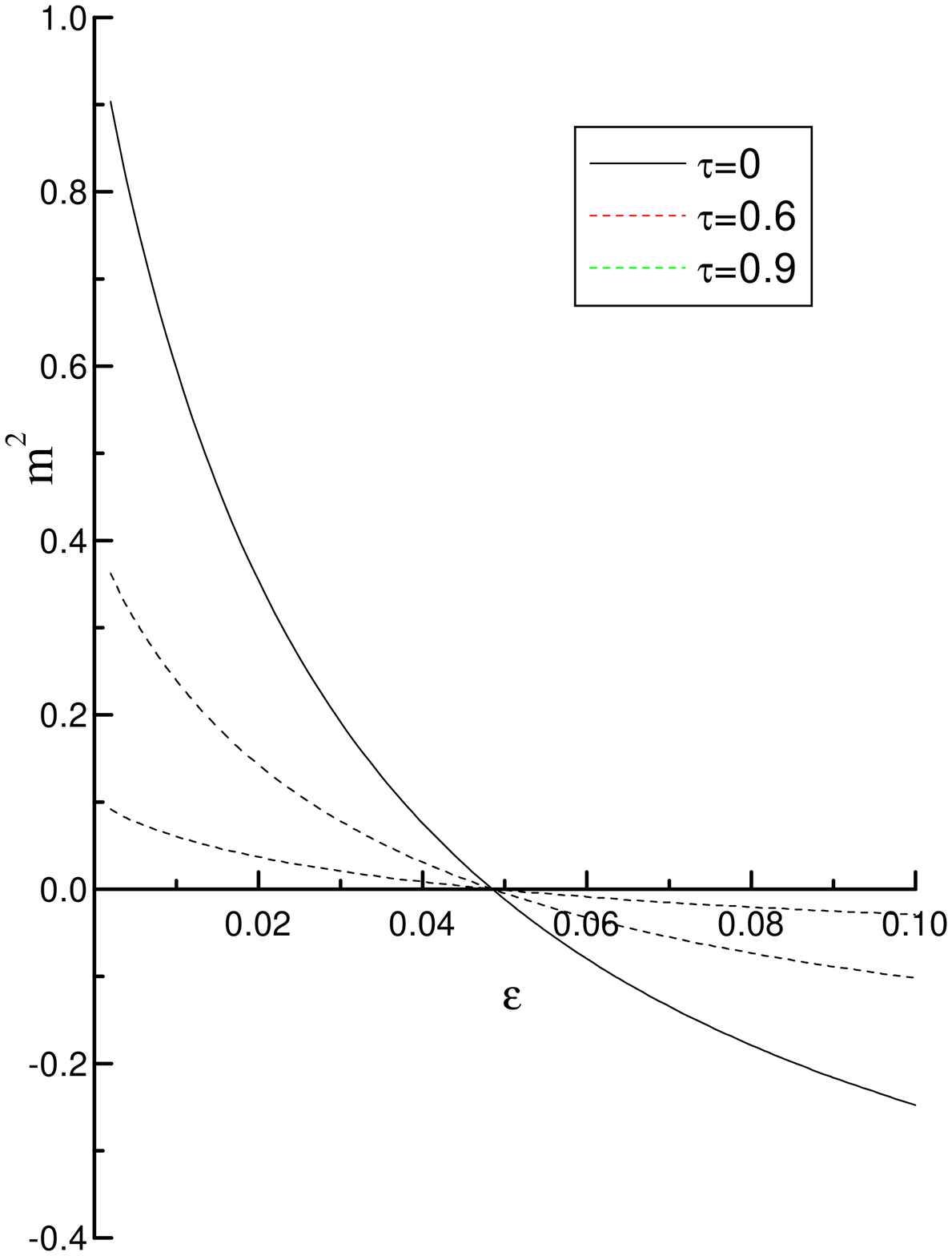}
 \end{center}
\caption{
The parameter  $m^2 $ of th GL model
 given by Eqs. \re{param} and \re{mas2dim} for $D=2+2\veps$.
}
\end{figure}
\newpage 
 \begin{figure}
 \epsfxsize=12.cm
\begin{center}
\epsffile{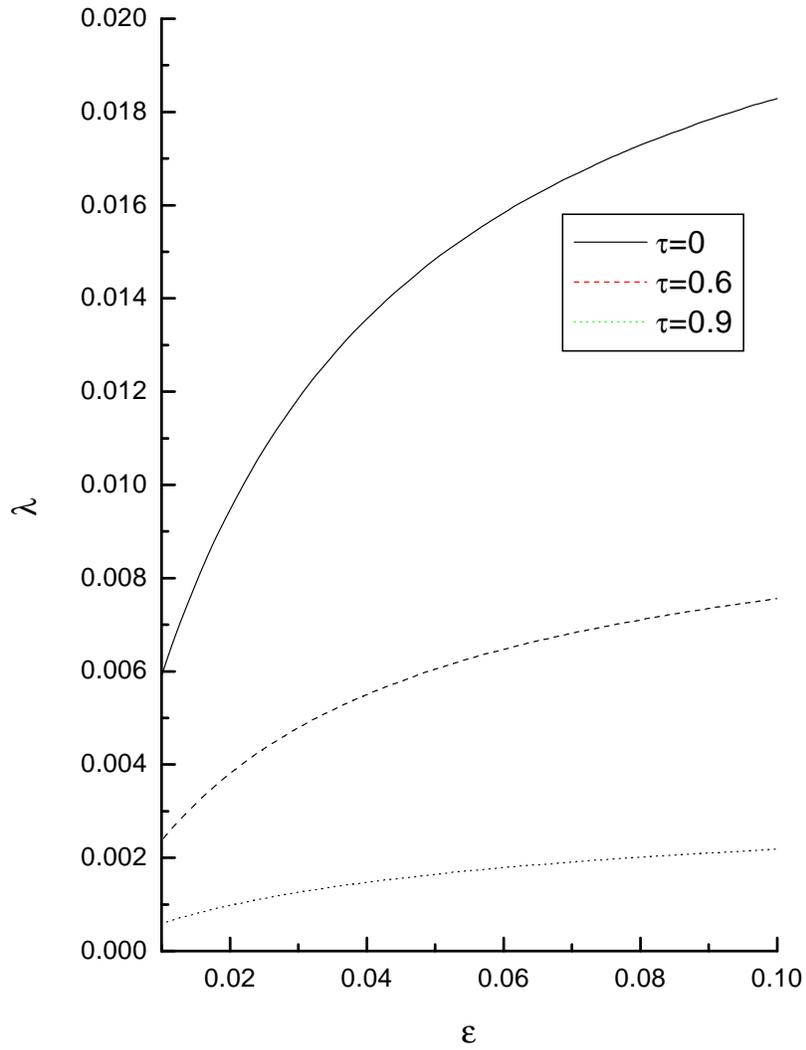}
 \end{center}
\caption{
The same as in Fig.3 but for $\lambda$
}
\end{figure}

\edc